\newcommand{\tTT}{t_{TT}}
\newcommand{\tTB}{t_{TB}}

\documentclass[aps,twocolumn,twocolumngrid,superbib]{revtex4}

\usepackage{graphicx}
\usepackage{amsfonts}
\usepackage{amsmath}
\usepackage{bm}
\usepackage{alltt}
\usepackage{dcolumn}
\usepackage{letterspace}


\begin{document}
\title[Short Title]{
Linear scaling computation of the Fock matrix. IX. \\
Parallel computation of the Coulomb matrix\footnotemark[1]}

\author{Chee Kwan Gan\footnotemark[2]}
\author{C. J. Tymczak}
\author{Matt Challacombe}

\affiliation{ Theoretical Division,\\ Los Alamos
              National Laboratory,\\ Los Alamos, New Mexico 87545}

\date{June 4, 2004}

\begin{abstract}
We present parallelization of a quantum-chemical tree-code
[J. Chem. Phys. {\bf 106}, 5526 (1997)] for linear scaling computation
of the Coulomb matrix.  Equal time partition [J. Chem. Phys. {\bf
118}, 9128 (2003)] is used to load balance computation of the Coulomb
matrix. Equal time partition is a measurement based algorithm for
domain decomposition that exploits small variation of the density
between self-consistent-field cycles to achieve load balance.
Efficiency of the equal time partition is illustrated by several tests
involving both finite and periodic systems.  It is found that equal time
partition is able to deliver 91 -- 98 \% efficiency with 128
processors in the most time consuming part of the
Coulomb matrix calculation.  The current parallel quantum chemical
tree code is able to deliver 63 -- 81\% overall efficiency on 128
processors with fine grained parallelism (less than two heavy atoms
per processor).



\end{abstract}
\maketitle

\footnotetext[1]{Preprint LA-UR-04-3626.}
\footnotetext[2]{ckgan@lanl.gov}

\section{Introduction}
\label{sec:intro}
Self-consistent-field (SCF) theories such as Density Functional Theory
and hybrid Hartree-Fock/Density Functional Theory are accurate and
computationally efficient. Traditional Gaussian-orbital quantum
chemistry codes that use conventional methods\cite{ASzabo89} are
usually restricted to small systems since these methods have steep
scaling of ${\cal O}(N^{2-3})$ with respect to system size, $N$.
Recently, significant progress has been made in the development of
${\cal O}(N)$ methods that overcome these bottlenecks.  These methods
include computation of the Hartree--Fock exchange matrix
\cite{ESchwegler96,ESchwegler97,ESchwegler98A,ESchwegler99,ESchwegler00,CTymczak04b},
the Coulomb
matrix~\cite{CWhite94B,CWhite96A,MChallacombe96,MChallacombe96B,MStrain96,JPerezjorda97,MChallacombe97,CTymczak04a},
the exchange-correlation
matrix~\cite{CTymczak04a,Jorda95,RStratmann96,CGuerra98,MChallacombe00A},
and iterative alternatives to eigensolution of the SCF
equations~\cite{XLi93,MDaw93,ADaniels97,APalser98,MChallacombe99,ANiklasson02A,ANiklasson03}.

With the advent of parallel multi-processor computers, especially
those based on commodity processors, there has been a great effort in
the community to parallelize quantum chemistry
codes\cite{MSchmidt93,MColvin93,Harrison_94v45,Guerra_95,TFurlani95,Sosa_98v19,vonArnim98,Furlani_00v128,Sosa_00v26,RKendall00,GFletcher00,Baker_02v23,JBaker04}.
Successful parallelization of ${\cal O}(N)$ methods hold promise for
large scale computations given the fact that with parallel linear
scaling methods, an $n$-fold increase in processors should lead to an
$n$-fold increase in simulation capability. However, this holds only
for scalable algorithms.

Two of the most computationally demanding parts in a density
functional application are calculation of the exchange-correlation and
Coulomb matrices. The ${\cal O}(N)$ exchange-correlation matrix
calculation has been efficiently parallelized through the concept of
equal time (ET) partition\cite{CGan03}.  In this work, the ET
partition is extended to load balancing calculation of the Coulomb
matrix.

Linear scaling computation of the Coulomb matrix has been achieved via
the quantum-chemical tree-code
(QCTC)\cite{MChallacombe96,MChallacombe96B,MChallacombe97} and the
continuous Fast Multipole Method
(CFMM)\cite{CWhite94B,CWhite96A,MStrain96}.  Both the
tree-code\cite{JBarnes86} and the Fast Multipole
Method\cite{LGreengard87,CRAnderson92} were originally proposed to
handle the astrophysical $N$-body problem.  Parallelization of these
$N$-body algorithms has been an active area of research in the
computer science
community\cite{MWarren92,AGrama94,MWarren95b,Singh93,Singh_95v27,YHu96,Grama_98v24,PGibbon02,Antonuccio-Delogu03}.
Even though both the $N$-body problem and the Coulomb matrix
calculation share many similarities, especially in handling the
far-field multipole contribution, little work has been done on the
parallel Coulomb matrix calculation beyond the simple master-slave
approach\cite{Sosa_98v19,Furlani_00v128,Sosa_00v26}.  It should be
pointed out that parallel ${\cal O}(N)$ computation of the Coulomb
matrix is highly irregular relative to parallel ${\cal O}(N^4)$
computation of the two-electron Coulomb integrals, where the jobs are
significantly coarse grained, enabling the master-slave approach to
work well.  It is well-known that the master-slave approach faces
potential contention and load imbalance problems for fine grained
parallelism\cite{BWilkinson99} (a small ratio of work load to number
of processors).  These problems have indeed been observed in quantum
chemical calculations\cite{Guerra_95,CGan03}, so alternatives are
needed.  One may use the idea of counting the number of interactions
in parallel $N$-body codes to load balance computation as in the
orthogonal recursive bisection (ORB)\cite{MWarren92} or Costzones
methods\cite{Singh93,Singh_95v27}.  However, due to cost
irregularities associated with different Gaussian extents, angular
symmetries, and non-uniform access patterns, simple counting is not an
optimal approach to load balance computation of the Coulomb matrix.

In this work, our main emphasis is on load balancing the most time
consuming part of QCTC, which is traversal of the density tree for
evaluation of Coulomb matrix elements. To load balance this highly
irregular tree traversal, we use the equal time (ET)
partition\cite{CGan03}, which was originally proposed to parallelize
computation of the exchange-correlation matrix.

Equal time partition works by measuring the time spent in
computational sub-domains ({\it e.g.} a line, area, or volume) during
one SCF cycle. At the end of the calculation, the time spent in each
sub-domain is used to predict a new overall domain decomposition for
the next SCF cycle, where each new sub-domain ideally incurs the same
amount of work in the next SCF cycle. The predicted domain
decomposition will deliver an improved load balance in the next SCF
cycle when there is a smooth variation of the workload between
successive SCF cycles ({\it e.g.} due to small changes in the electron
density).  In this way, temporal locality\cite{JPilkington96} of the
problem is exploited to achieve a continuously improved load balance.

In serial, the time to build the density tree constitutes about 2\% or
less of the total time spent in QCTC.  Unfortunately, Amdahl's law
dictates that the performance of a massively parallel program is
ultimately determined by its serial parts.  Therefore we also need to
consider parallel construction of the density tree. Again, ideas from
parallel $N$-body codes may be useful.  The construction of locally
essential trees, which are {\it just} sufficient for tree traversal on
each processor,\cite{MWarren92} avoids the problem of replicating the
total density tree on each processor.  Hashed
oct-trees\cite{MWarren93,MWarren95b} also solve the replication
problem, where hash tables are used to allow the program to access
data in an efficient manner across multiple processors. However, due
to fact that these approaches entail significant code restructuring,
for the present case, we have chosen a parallel replicated density
tree approach.

The remainder of this paper is organized as follows: In
Section~\ref{ParaQCTC} we discuss our strategy to efficiently
parallelize computation of the Coulomb matrix ${\bf J}$. In
Section~\ref{sec:implementation} we describe a computational
implementation of parallel QCTC. In Section~\ref{results} we discuss
results of speedup tests performed on a few representative finite and
periodic systems. In Section~\ref{conclusions} we summarize the main
conclusions of the paper.

\section{Parallelization of QCTC}
\label{ParaQCTC}
The quantum-chemical tree-code (QCTC) for ${\cal O}(N)$ calculation of
the Coulomb matrix has been fully described in
Refs.~[\onlinecite{MChallacombe97}] and [\onlinecite{CTymczak04a}].
Here we only highlight essential aspects of the algorithm so that
discussions of parallelization of QCTC may be made.

The Coulomb matrix element in a finite (gas phase) case is given by
\begin{equation}
{\bf J}_{ab} = \int d{\bf r}d{\bf r}' \frac{\rho_{ab}({\bf r})
\rho_{\rm tot}({\bf r}')}{| {\bf r} - {\bf r}'|},
\label{eq:Jab}
\end{equation}
where the charge distribution\cite{LMcmurchie78} (or simply the
distribution) $\rho_{ab}({\bf r})$, is a product of the (Gaussian)
basis functions $\phi_a({\bf r})$ and $\phi_b({\bf r}) $.  The total
density of the system, which includes both the electronic and nuclear
parts, is denoted by $\rho_{\rm tot}({\bf r})$.  In QCTC, a
hierarchical multipole representation of the electron density, called
a density tree, is stored in an advanced $k$-d tree data
structure\cite{Bentley79,Bentley80,Gaede98}.  A compact representation
of the density in terms of Hermite-Gaussian
(HG)\cite{MChallacombe97,MChallacombe00A,GAhmadi95} basis has been
used.  With the help of the density tree, QCTC re-expresses the matrix
element in Eq.~(\ref{eq:Jab}) as a sum of near-field (NF) and
far-field (FF) terms\cite{CTymczak04a}
\begin{eqnarray}
J_{ab} &=& \sum_{Q\in \rm FF} \sum_{\ell} (-1)^{\ell} \sum_{m}
O^\ell_m[\rho_{ab}]
\sum_{\ell'} \sum_{m'} M^{\ell+\ell'}_{m+m'} O^{\ell'}_{m'}[\rho_Q]
\nonumber \\
&+& \sum_{q\in \rm NF} \int d {\bf r} \int d {\bf r'} \rho_{ab} ({\bf
r}) \left|{\bf r}-{\bf r'} \right|^{-1}
\rho_q({\bf r})
\label{eq:JabTree}
\end{eqnarray}
where $M^\ell_m$ is the irregular solid harmonic interaction tensor,
$O^\ell_m[f]=\int d{\bf r} O^\ell_m({\bf r}) f({\bf r})$ is a moment
of the regular solid harmonics, $Q$ runs over the all nodes in the
density tree as determined by penetration admissibility criterion
(PAC) and multipole admissibility criterion
(MAC)\cite{MChallacombe97,CTymczak04a} and $q$ runs on the left over
near-field primitive distributions in the density. For the periodic
case, a periodic far-field term and a tin-foil boundary condition
term\cite{MChallacombe97D,CTymczak04a} are added to the RHS of
Eq.~(\ref{eq:JabTree}).

Essential to QCTC is construction of the total density tree.  Once the
density tree is built, calculation of the Coulomb matrix elements
proceeds by transversing the tree and checking the PAC and MAC. When
both PAC and MAC are met, the far-field contribution is calculated via
the multipole approximation.  The near-field contribution, however, is
calculated analytically.

From Eq.~(\ref{eq:JabTree}), it is easy to see that ${\cal O}(N)$
computation of the Coulomb matrix is highly irregular; near- and
far-field contributions are determined on the fly via a PAC and MAC
that depends on both the distributions and the density. This poses a
challenge to efficiently load balance parallel tree traversal in QCTC.

\subsection{Load balancing tree traversals}
\label{ETPartition}
Since ET partition involves measuring workload using a timer (e.g. the
{\tt MPI\_WTime} function in the message-passing libary
MPI\cite{mpi}), it is important to decide what work load information
to time so that the timing process itself does not incur too much
overhead.  This may be achieved by recording the time to traverse the
tree for each distribution (i.e. $\rho_{ab}({\bf r})$ in
Eq.~(\ref{eq:Jab})). The time and position of each distribution are
stored in an array on each processor to facilitate partitioning of the
3-D bounding box (the root bounding box) that encloses all
distributions. Equal time partition\cite{CGan03} is performed on the
root bounding box to achieve equal time or cost in all sub-boxes (also
called ET sub-boxes).  ET partition creates ET sub-boxes by
recursively partitioning a box into 2 sub-boxes such that each sub-box
carries approximately the same amount of time. At the end of the
procedure we obtain $2^n$ ET sub-boxes, where $n$ is an integer
greater than zero.  Assuming that the number of processors is $2^n$,
each processor will handle one of the $2^n$ ET sub-boxes in the next
SCF cycle.  The restriction of a power of two for the number of
processors may be removed by using a general ET partitioning scheme
detailed in the Appendix of Ref.[\onlinecite{CGan03}].  We have used a
robust bisection method\cite{WPress92} to find the plane which
approximately divides the workload into half.  We emphasize again the
main difference between our ET scheme and other parallel $N$-body
codes\cite{MWarren92,Singh93} is that we use an exact load timing
information rather than counting the number of interactions as in the
orthogonal recursive bisection (ORB)\cite{MWarren92} and Costzones
methods\cite{Singh93}.

For the periodic case, the time associated with each distribution
includes the time to handle the periodic far
field\cite{MChallacombe97D,CTymczak04a} contribution. In this way, ET
partition naturally uses combined timing information to load balance
computation, extending the power of ET partition to situations where
different timing information may be grouped together.

A working hypothesis of the ET partition applied to QCTC is that the
sum of distribution times for each ET sub-box is constant irrespective
of the sectioning. However, for very fine grained parallelism,
shifting a bisecting plane may induce a relatively large change in the
total predicted distribution time in a sub-box. This is due to the
fact that the total workload may not be equally divided among the
sub-boxes because the distribution times are discrete (a distribution
is either {\it totally} or not in a sub-box).  Also, for very fine
grained parallelism, the total work in a sub-box is more sensitive to
a change in density that may also increase the load imbalance, an
effect which we have experienced in parallelization of the
exchange-correlation matrix\cite{CGan03}.

In the first cycle, there is no previous timing on which to base the
ET. In such a case, we use a reasonable heuristic where each processor
handles an approximately equal number of distributions (the total
number of distributions may not be exactly divisible by the number of
processors).

\subsection{Parallel density tree build} 
\label{sec:parallelTB}

In principle, an efficient parallelization of the density tree build
should make use of the fact that each processor is handling only part
of the total distributions in an ET sub-box.  Depending on the
collection of distributions on each processor, a locally essential
tree\cite{MWarren92,CGan03} may be constructed which is just sufficient for the tree
traversals of all the distributions on a
processor, thus avoiding replication of the
entire density tree and enabling efficient use of the memory space.
However, without an extensive programming task, it may be difficult to
predetermine which part of the entire density tree is needed for
construction of the locally essential tree. As a first attempt, we
have chosen a simple approach to parallelize the entire density tree
build.

For simplicity of programming we assume the number of processors to be
$2^k$, where $k$ is an integer greater than zero.  Observing that
there are $2^k$ subtrees at the $k$th tier in the entire density tree,
our current implementation adopts a simple scheme where each processor
builds one of the $k$th-tier subtrees in the total density tree. When
all processors have built a $k$th-tier subtree, an all-to-all exchange
is carried out where all processors get the rest of the $k$th-tier
subtrees so that a final ``merging'' up of the
subtrees\cite{MChallacombe97,CTymczak04a} can be performed to obtain
the entire density tree.

Inefficiency of the current implementation of the parallel density
tree build in the limit of a large number of processors is expected.
The all-to-all exchange of data between processors is expensive and
does not scale with the number of processors. Also, after collecting
the subtrees from all other processors, a processor has to ``merge''
more subtrees upward as the number of processors is increased. This
will inevitably introduce more overhead as we use more processors.
Even if one can overcome the all-to-all exchange problem, one still
faces a problem where it may be wasteful in the use of memory to store
the entire density tree on each processor.  However, while the present
parallel density tree build may be replaced by more sophisticated
schemes, where locally essential trees are built\cite{MWarren92} or
hashed trees are used\cite{MWarren93,MWarren95b}, the current
implementation delivers very good speedups up to the 128-processor
level.

As a side note, our first implementation of an ET parallel QCTC tried
to avoid the problem mentioned above by partitioning the entire
density into disjoint local densities. Each processor then built a
local tree based on the local density. However, since a distribution
on one processor does not ``see'' other local density trees, every
processor had to loop through all distributions and the resulting
partial Coulomb matrices had to be resummed using an all-to-all
communication at the end of the calculation. This turned out to be
practical only below the 64-processor level. The speedup did not
increase with more processors because the total intrinsic cost
(i.e. the amount of useful work) of QCTC {\it increases} rather
rapidly with the number of processors. The rapid increase of intrinsic
cost has at its root a break down of the hierarchical multipole
approximation, as physically close charges can no longer be grouped
when they reside on different processors.  Asymptotically, as the
number of processors approach the number of charges, one reverts to
the expensive ${\cal O}(N^2)$ algorithm.  For periodic systems, where
one has to visit the density tree many more times (looping through
periodic images) relative to the finite case, the speedup stagnates
once we pass a certain number of processors. Since this version of the
parallel QCTC does not scale with the number of processors, we do not
consider it further in this work.

\section{Implementation}
\label{sec:implementation}
We have implemented a parallel QCTC algorithm in {\sc
MondoSCF}~\cite{MondoSCF}, a suite of programs for linear scaling
electronic structure theory and {\it ab initio}\/ molecular dynamics.
{\sc MondoSCF} has been written in Fortran 90/95 with the
message-passing library MPI~\cite{mpi}.  Timings are performed using
the {\tt MPI\_WTIME} function.

\section{Results}
\label{results}
We have performed scaling tests on both finite and periodic
systems. For the finite systems, we have chosen taxol
(C$_{47}$H$_{51}$NO$_{14}$) and 2 water clusters as test cases. For
the periodic systems, we have chosen pentaerythritol tetranitrate
(PETN)\cite{CGan04A} and the $\delta$-phase of
octahydro-1,3,5,7-tetranitro-1,3,5,7-tetrazocine
($\delta$-HMX)\cite{JPLewis00} as representative test cases. These
systems are chosen because they are highly inhomogeneous,
three-dimensional (and two are periodic) systems posing a challenge to
parallel QCTC. All runs were performed on a cluster of 256 4 CPU
HP/Compaq Alphaserver ES45s with the Quadrics QsNet High Speed
Interconnect.

For the purpose of performing the scaling tests, we start the
calculation with the STO-3G basis set and a low accuracy, and switch
to the final basis set and accuracy using a mixed integral approach,
and run for three SCF cycles. The density matrix ${\bf P}$ is saved to
disk and scaling tests of parallel QCTC are performed. This procedure
may not be necessary. However, we are confident that the timings are
representatives of a routine calculation.

\begin{figure}[t]
\resizebox*{3.5in}{!}{\includegraphics[clip]{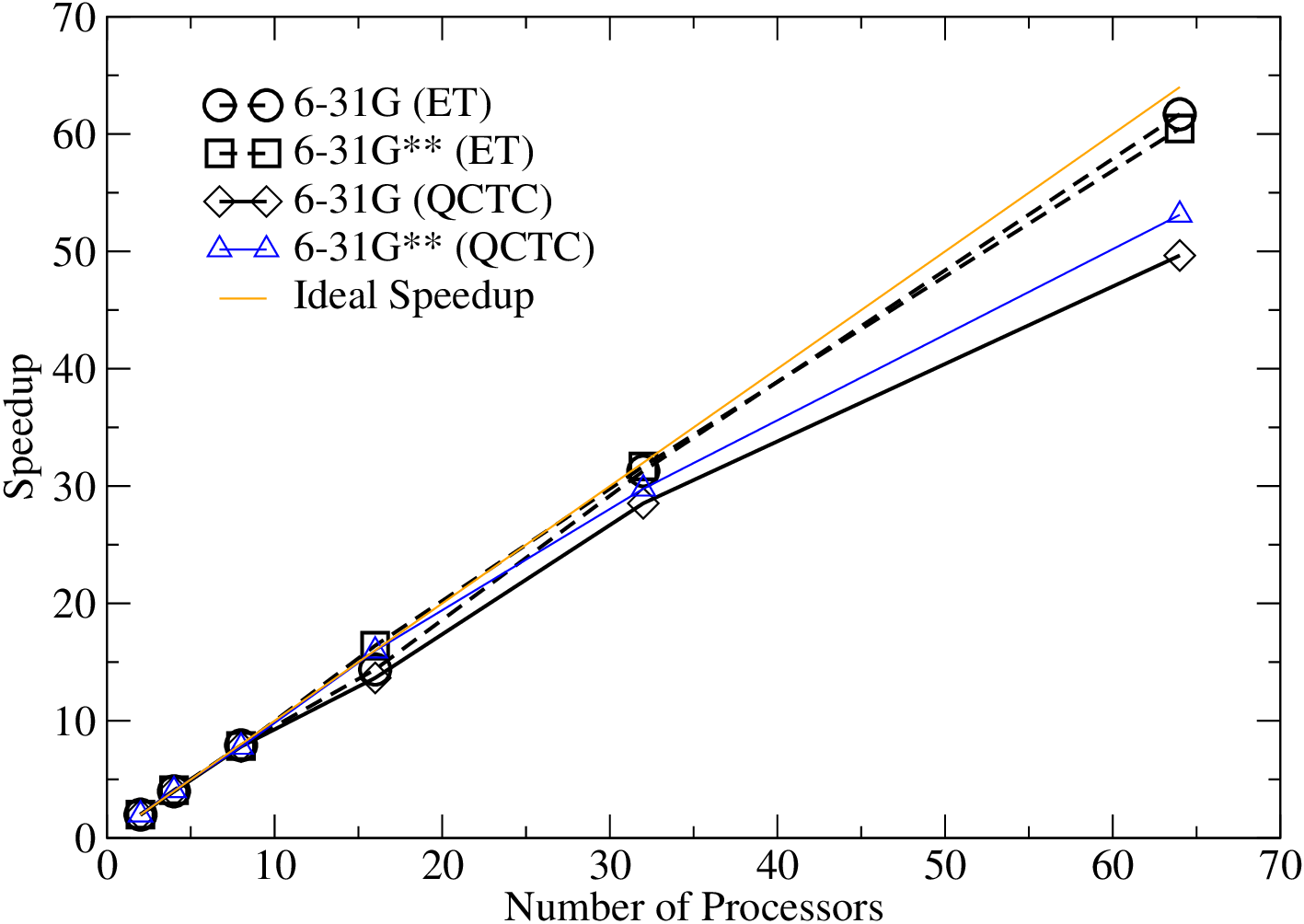}}
\caption{ 
Scaling of parallel QCTC on taxol (C$_{47}$H$_{51}$NO$_{14}$)
RBLYP/6-31G and RBLYP/6-31G**. Speedups are relative to a 2-processor
calculation.  The labels ET and QCTC denote equal time and overall
parallel QCTC speedups, respectively.  }
\label{fig:taxol}
\end{figure}

The result of the taxol scaling test is shown in Fig.~\ref{fig:taxol}.
The calculations are performed with the 6-31G and 6-31G** basis sets,
and a {\tt GOOD} accuracy\cite{CTymczak04a}.  The results of two
different speedups are presented.  The first speedup, called the ET
speedup, measures the efficiency of the ET partition for the Coulomb
matrix element calculation by traversing the density tree (see
Section~\ref{ETPartition}) and is defined by
\begin{equation}
S_{ET} = \frac{2\tTT^{(2)}}{\tTT^{(n)}}
\end{equation}
where $\tTT^{(n)}$ is the time to evaluate the matrix elements by
traversing the density tree with $n$ processors. Notice that the
speedups are relative to a 2-processor calculation.  The second
speedup, called the QCTC speedup, measures the overall efficiency of
parallel QCTC.  From Fig.~\ref{fig:taxol} it is observed that the ET
speedup is excellent up to 64 processors. Efficiency at the
64-processor level (where the number of heavy atoms per processor is
less than 1) is at least 94\%.
The overall parallel QCTC speedup is very good up to 32 processors but
degrades slightly at the 64-processor level.  The overall parallel
QCTC efficiencies at 64 processors are
77.6\% and 83.0\% for the 6-31G and 6-31G** basis sets, respectively.

\begin{figure}[t]
\resizebox*{3.5in}{!}{\includegraphics[clip]{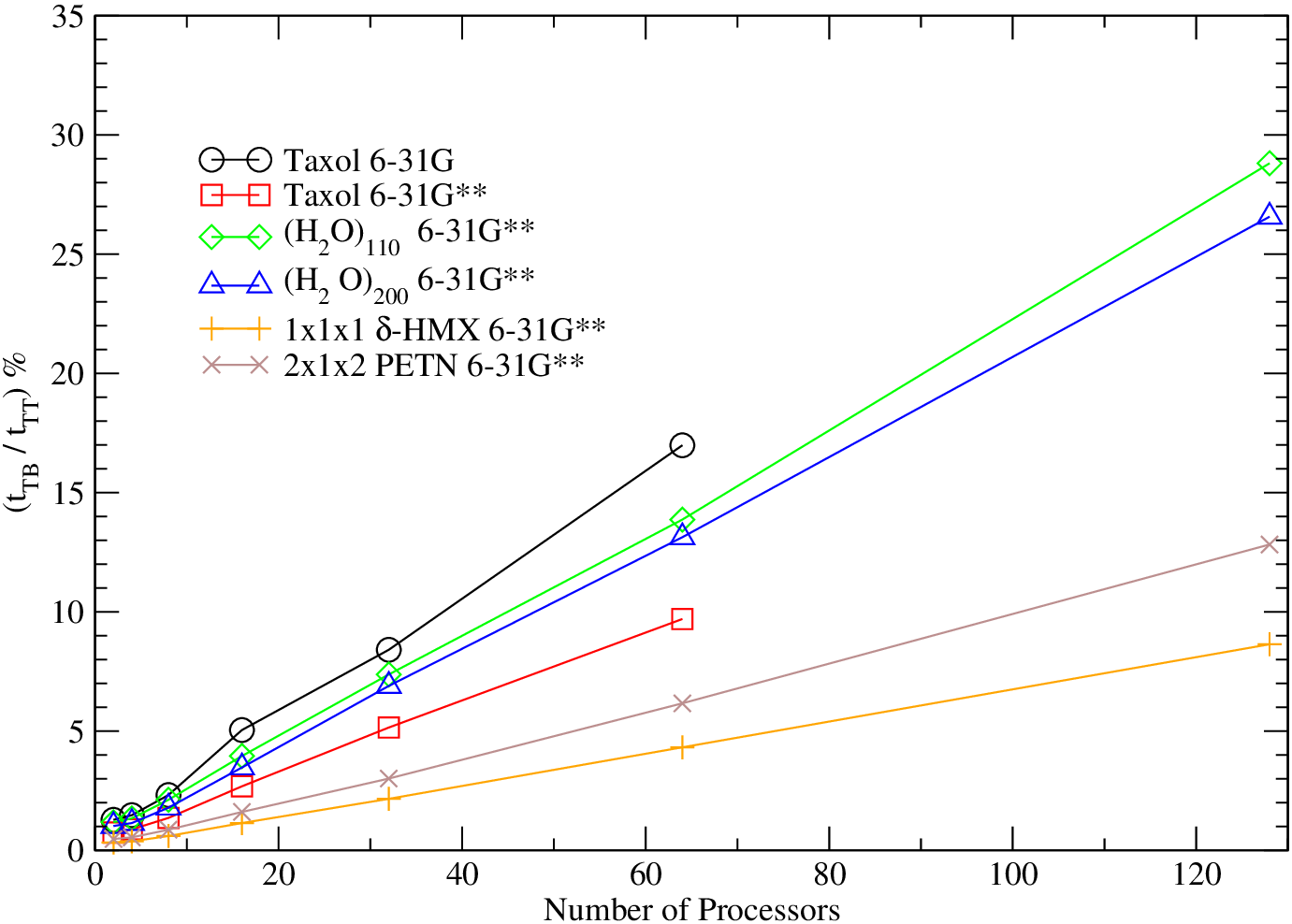}}
\caption{ 
The ratio of the time to build the density tree ($\tTB$) to the time
to traverse the density tree ($\tTT$), as a function of the number of
processors.  }
\label{fig:time_ratio}
\end{figure}

The loss of efficiency is due to the fact that the time for parallel
density tree build does not decrease at the same rate (i.e.  divided
by the number of processors) as the tree traversal part.  In
Fig.~\ref{fig:time_ratio} we present the $\tTB/\tTT$ ratio as a
function of the number of processors for calculations on taxol (along
with other systems for comparisons to be made later).  We note that if
the time for parallel tree build $\tTB$ were to decrease at the same
rate (i.e. divided by the number of processors) as the time for tree
traversal $\tTT$ as we increase the number of processors, then the
$\tTB/\tTT$ ratio would remain nearly constant.  However,
Fig.~\ref{fig:time_ratio} shows that the $\tTB/\tTT$ ratio increases
steadily as the number of processors is increased in all cases, a fact
that has been anticipated from the discussion in
Section~\ref{sec:parallelTB}.  Since the slope for the 6-31G** case is
smaller than that for the 6-31G case, this explains the slight
increase in the overall parallel QCTC performance of the 6-31G** case
over the 6-31G case, as shown in Fig.~\ref{fig:taxol}.

We note that our results of a parallel QCTC
speedup of 7.80 (with the 6-31G and 6-31G** basis sets) with 8
processors compares favorably with the speedup of about 6.0 of Sosa
{\it et al.}\cite{Sosa_00v26}, which is for an entire single-point
energy calculation with RHF/3-21G.

\begin{figure}[t]
\resizebox*{3.5in}{!}{\includegraphics[clip]{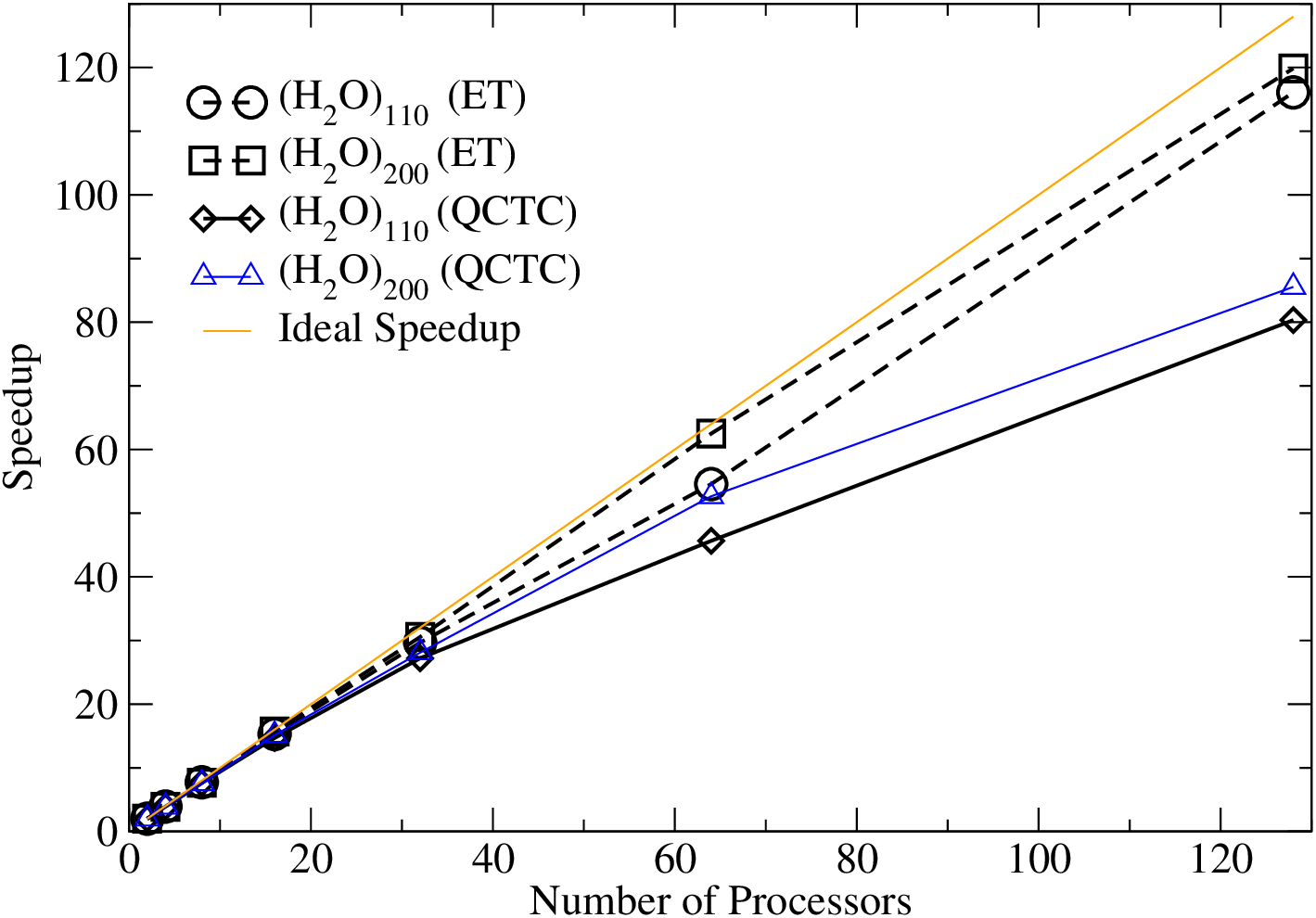}}
\caption{ 
Scaling of parallel QCTC on 2 cluster of water molecules with
RBLYP/6-31G**. Speedups are relative to a 2-processor calculation.  }
\label{110And200WaterOnGood}
\end{figure}

\begin{figure}[t]
\resizebox*{3.5in}{!}{\includegraphics[clip]{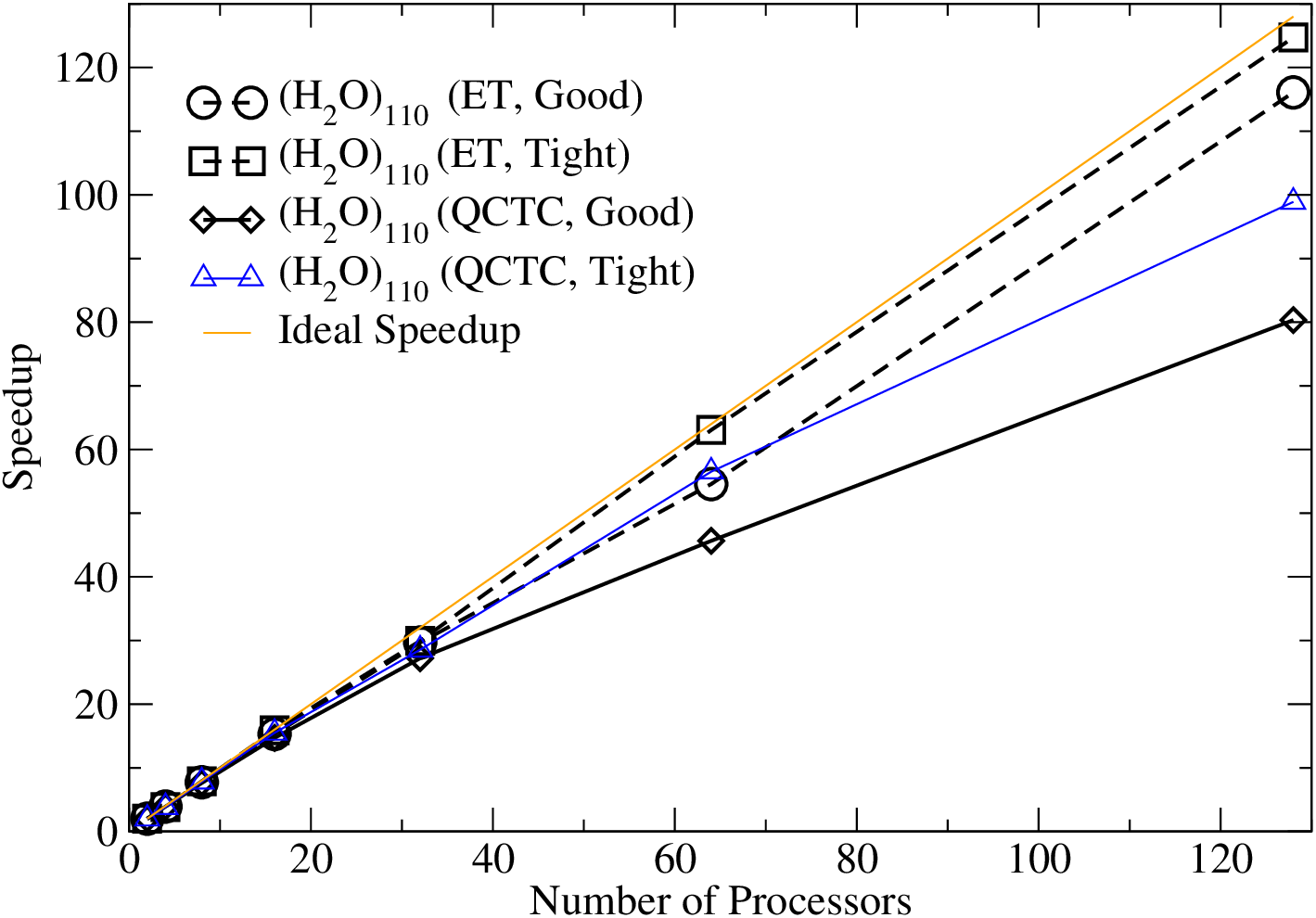}}
\caption{ 
Scaling of parallel QCTC on 110-molecule water cluster with
RBLYP/6-31G** on {\tt GOOD} and {\tt TIGHT} accuracies.  Speedups are
relative to a 2-processor calculation.  }
\label{fig:110WaterOnGoodAndTight}
\end{figure}

\begin{figure}[t]
\resizebox*{3.5in}{!}{\includegraphics[clip]{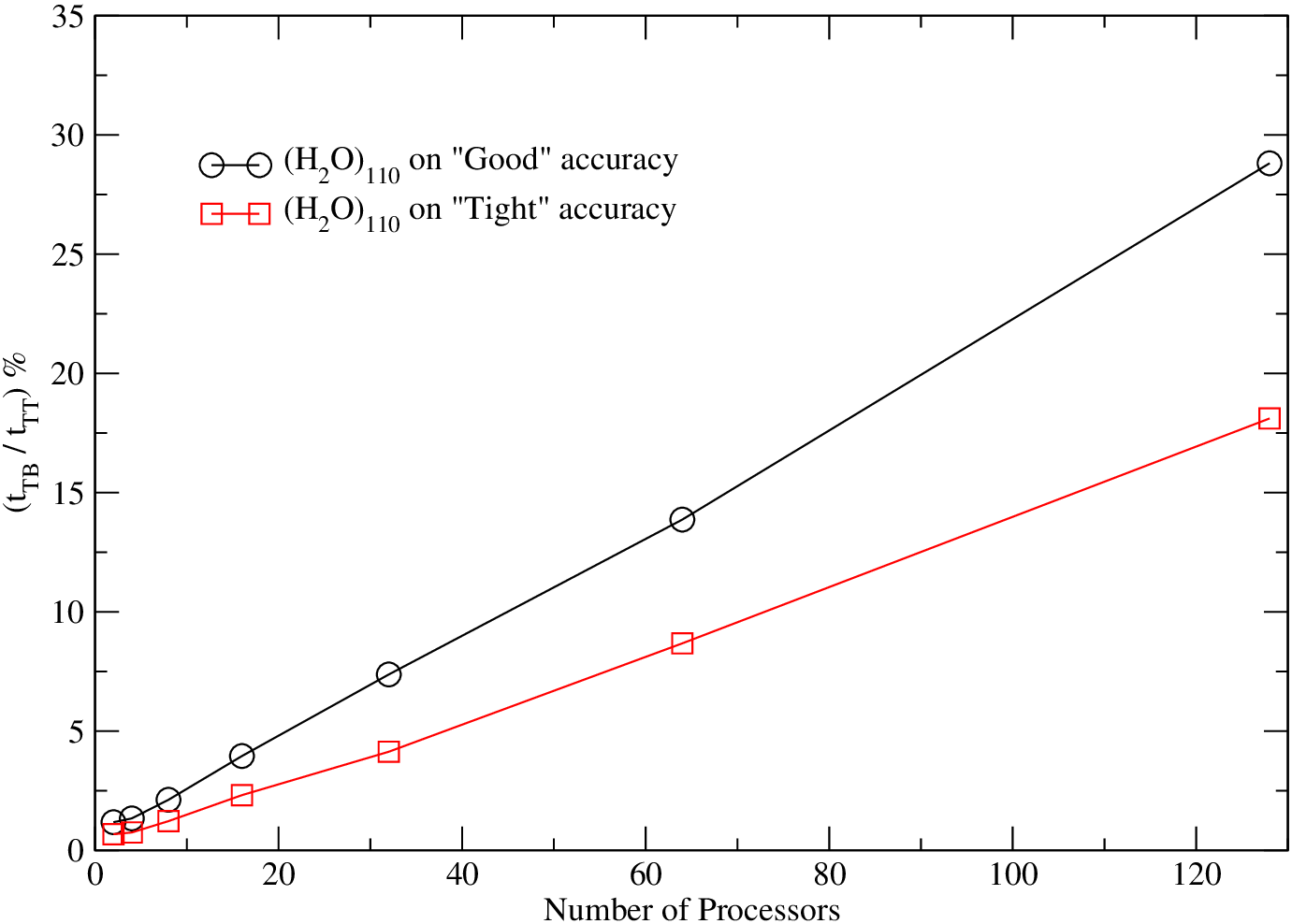}}
\caption{ 
The ratio of the time to build the density tree ($\tTB$) to the the
time to traverse the density tree ($\tTT$), as a function of the
number of processors, for 110-molecule water cluster (RBLYP/6-31G**)
calculations on {\tt GOOD} and {\tt TIGHT} accuracies.}
\label{fig:NProc_TRatio_110H2O_GoodAndTight}
\end{figure}

Similar scaling tests have been performed on a 110-molecule and
200-molecule water clusters with RBLYP/6-31G** at a {\tt GOOD}
accuracy level.  The result of the scaling tests is shown in
Fig.~\ref{110And200WaterOnGood}. It is found that the ET speedups are
rather good for both cases. The overall parallel QCTC speedups are
80.3 and 85.5 for the 128-processor calculations for 110-molecule and
200-molecule water clusters, respectively.
The decrease in parallel QCTC efficiency is again due to the
high $\tTB/\tTT$ ratio at the 128-processor level. These ratios are
28.8\% and 26.6\% for the 110-molecule and 200-molecule water
clusters, respectively (see Fig.~\ref{fig:time_ratio}).

To investigate the performance of parallel QCTC at a higher accuracy
level, we have performed the scaling tests on a 110-molecule water
cluster but with {\tt TIGHT} accuracy\cite{CTymczak04a}. The results for both {\tt GOOD}
and {\tt TIGHT} accuracies are presented in
Fig.~\ref{fig:110WaterOnGoodAndTight} for comparison.  It is seen that
the ET speedup is better for the {\tt TIGHT} case than for the {\tt
GOOD} case, which is anticipated since increasing the accuracy level
increases the effective granularity, which leads to a better
performance in ET partition\cite{CGan03}.  Overall parallel QCTC
increases its efficiency from {\tt GOOD} to {\tt TIGHT}, which is due
mainly to a decrease in the $\tTB/\tTT$ ratio, as shown in
Fig.~\ref{fig:NProc_TRatio_110H2O_GoodAndTight}.

\begin{figure}[t]
\resizebox*{3.5in}{!}{\includegraphics[clip]{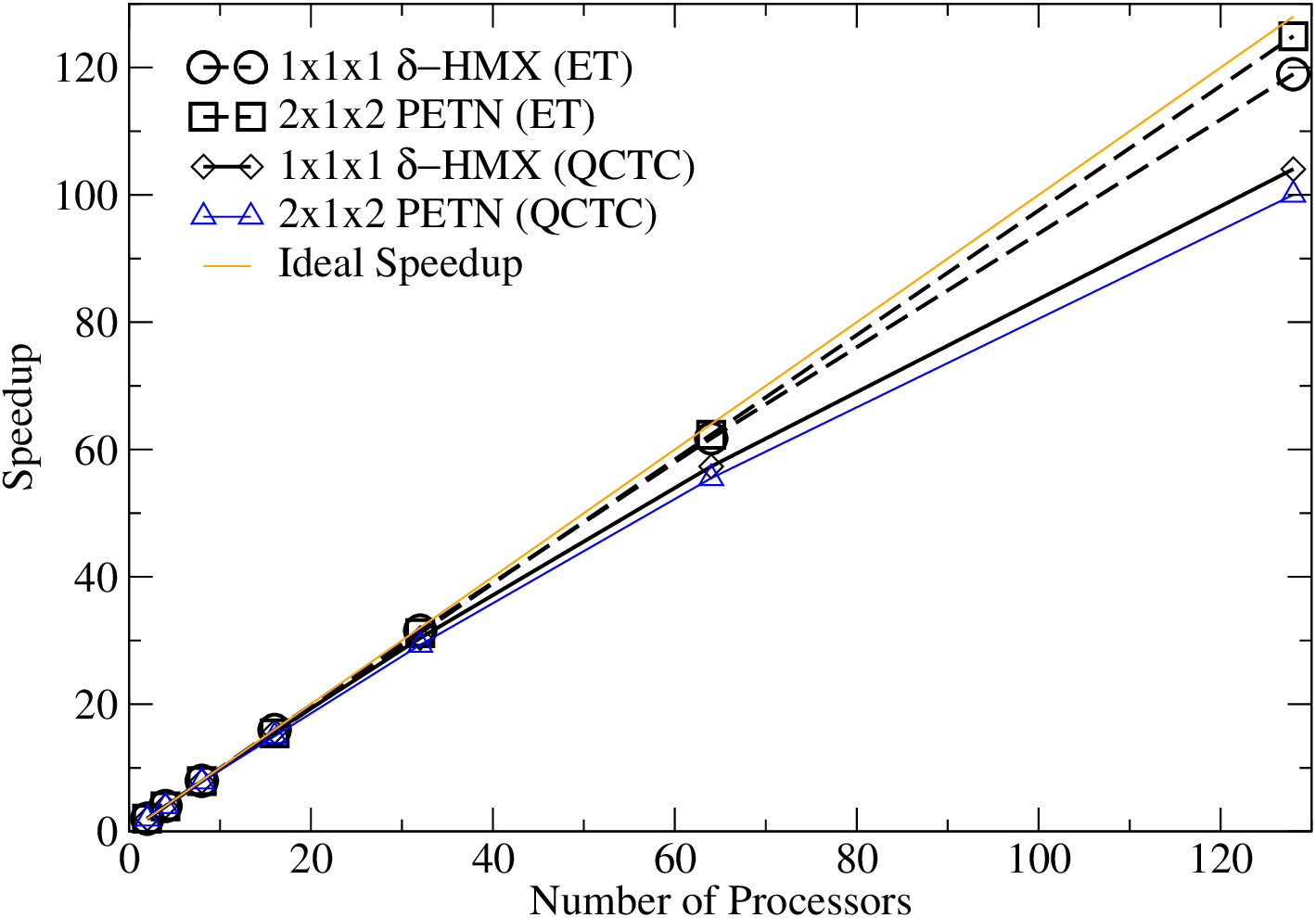}}
\caption{ 
Scaling of parallel QCTC on $\delta$-HMX and PETN with RPBE/6-31G**.
Speedups are relative to a 2-processor calculation.  }
\label{fig:111DHMX212PETN}
\end{figure}

Finally, for the periodic systems Fig.~\ref{fig:111DHMX212PETN} shows
that the overall parallel QCTC with RPBE/6-31G** on {\tt GOOD}
accuracy is excellent. At the 128-processor level, the $1\times
1\times 1$ $\delta$-HMX (168 atoms per simulation cell) delivers 104.0
fold speedup, while the $2\times 1 \times 2$ PETN (232 atoms per
simulation cell) delivers 100.0 fold speedup. These performances are
better compared to the 110-molecule (a speedup of 80.3) or
200-molecule (a speedup of 85.5) water cluster calculations. This is
due to the smaller $\tTB/\tTT$ ratio (see Fig.~\ref{fig:time_ratio})
for the periodic cases compared to the finite cases, which mainly
results from the increase in the time spent in the tree traversal part
(e.g. 87.7 and 64.0 secs for the $2\times 1 \times 2$ PETN and
200-molecule water cluster, respectively).

\section{Conclusions}
\label{conclusions}
We have proposed an efficient method of parallelizing calculation of
the Coulomb matrix. The concept of equal time (ET) has proven fruitful
for load balancing the most time consuming part of QCTC, which is
traversal of the density tree for the matrix element calculation.
Equal time exploits the temporal locality between SCF iterations to
overcome strong spatial irregularities. It is expected that ET should
retain this property between geometry steps in an optimization or
molecular dynamics run.  The efficiency of the ET partition ranges
from 91 -- 98 \% for all test cases presented in this work at the
128-processor level. The overall QCTC speedup, however, ranges from 63
-- 81 \% overall efficiency at the 128-processor level with fine
grained parallelism.  The decrease in efficiency is mainly due to the
parallel tree build process.  While the current simple implementation
of the parallel tree build should eventually be replaced by a more
sophisticated version, the current implementation has enabled us to
run routine calculations to address a wide range of interesting
problems\cite{CGan04C,CGan04A}.

\begin{acknowledgments}
This work has been carried out under the auspices of the
U.S. Department of Energy under Contract No.~W-7405-ENG-36 and the
ASCI project.  Most work was performed on the computing resources at
the Advanced Computing Laboratory of Los Alamos National Laboratory.
\end{acknowledgments}


\begin{thebibliography}{67}
\expandafter\ifx\csname natexlab\endcsname\relax\def\natexlab#1{#1}\fi
\expandafter\ifx\csname bibnamefont\endcsname\relax
  \def\bibnamefont#1{#1}\fi
\expandafter\ifx\csname bibfnamefont\endcsname\relax
  \def\bibfnamefont#1{#1}\fi
\expandafter\ifx\csname url\endcsname\relax
  \def\url#1{\texttt{#1}}\fi
\expandafter\ifx\csname urlprefix\endcsname\relax\def\urlprefix{URL }\fi
\providecommand*{\bibinfo}[2]{#2}
\providecommand*{\eprint}[1]{#1}
\providecommand*{\url}[1]{#1}
\begingroup\makeatletter
 \@temptokena{%
  \expandafter\ifx\csname citenamefont\endcsname\relax
   \DeclareRobustCommand\citenamefont{\@firstofone}%
   \global\let\citenamefont\citenamefont
   \global\expandafter\let\csname citenamefont \expandafter\endcsname\csname
  citenamefont \endcsname
  \fi
 }\if@filesw\immediate\write\@auxout{\the\@temptokena}\fi
\expandafter\endgroup\the\@temptokena

\bibitem[{\citenamefont{Szabo and Ostlund}(1989)}]{ASzabo89}
\bibinfo{author}{\bibfnamefont{A.}~\bibnamefont{Szabo}} \bibnamefont{and}
  \bibinfo{author}{\bibfnamefont{N.~S.} \bibnamefont{Ostlund}},
  \emph{\bibinfo{title}{Modern Quantum Chemistry}} (\bibinfo{publisher}{Mc
  Graw--Hill Inc.}, \bibinfo{address}{New York}, \bibinfo{year}{1989}), first,
  revised ed.

\bibitem[{\citenamefont{Schwegler and Challacombe}(1996)}]{ESchwegler96}
\bibinfo{author}{\bibfnamefont{E.}~\bibnamefont{Schwegler}} \bibnamefont{and}
  \bibinfo{author}{\bibfnamefont{M.}~\bibnamefont{Challacombe}},
  \bibinfo{journal}{J. Chem. Phys.} \textbf{\bibinfo{volume}{105}},
  \bibinfo{pages}{2726} (\bibinfo{year}{1996}).

\bibitem[{\citenamefont{Schwegler} \emph{et~al.}(1997)\citenamefont{Schwegler,
  Challacombe, and Head-Gordon}}]{ESchwegler97}
\bibinfo{author}{\bibfnamefont{E.}~\bibnamefont{Schwegler}},
  \bibinfo{author}{\bibfnamefont{M.}~\bibnamefont{Challacombe}},
  \bibnamefont{and}
  \bibinfo{author}{\bibfnamefont{M.}~\bibnamefont{Head-Gordon}},
  \bibinfo{journal}{J. Chem. Phys.} \textbf{\bibinfo{volume}{106}},
  \bibinfo{pages}{9708} (\bibinfo{year}{1997}).

\bibitem[{\citenamefont{Schwegler} \emph{et~al.}(1998)\citenamefont{Schwegler,
  Challacombe, and \mbox{Head-Gordon}}}]{ESchwegler98A}
\bibinfo{author}{\bibfnamefont{E.}~\bibnamefont{Schwegler}},
  \bibinfo{author}{\bibfnamefont{M.}~\bibnamefont{Challacombe}},
  \bibnamefont{and}
  \bibinfo{author}{\bibfnamefont{M.}~\bibnamefont{\mbox{Head-Gordon}}},
  \bibinfo{journal}{J. Chem. Phys.} \textbf{\bibinfo{volume}{109}},
  \bibinfo{pages}{8764} (\bibinfo{year}{1998}).

\bibitem[{\citenamefont{Schwegler and Challacombe}(1999)}]{ESchwegler99}
\bibinfo{author}{\bibfnamefont{E.}~\bibnamefont{Schwegler}} \bibnamefont{and}
  \bibinfo{author}{\bibfnamefont{M.}~\bibnamefont{Challacombe}},
  \bibinfo{journal}{J. Chem. Phys.} \textbf{\bibinfo{volume}{111}},
  \bibinfo{pages}{6223} (\bibinfo{year}{1999}).

\bibitem[{\citenamefont{Schwegler and Challacombe}(2000)}]{ESchwegler00}
\bibinfo{author}{\bibfnamefont{E.}~\bibnamefont{Schwegler}} \bibnamefont{and}
  \bibinfo{author}{\bibfnamefont{M.}~\bibnamefont{Challacombe}},
  \bibinfo{journal}{Theor. Chem. Acc.} \textbf{\bibinfo{volume}{104}},
  \bibinfo{pages}{344} (\bibinfo{year}{2000}).

\bibitem[{\citenamefont{Tymczak} \emph{et~al.}(2004)\citenamefont{Tymczak,
  Weber, Schwegler, and Challacombe}}]{CTymczak04b}
\bibinfo{author}{\bibfnamefont{C.~J.} \bibnamefont{Tymczak}},
  \bibinfo{author}{\bibfnamefont{V.}~\bibnamefont{Weber}},
  \bibinfo{author}{\bibfnamefont{E.}~\bibnamefont{Schwegler}},
  \bibnamefont{and}
  \bibinfo{author}{\bibfnamefont{M.}~\bibnamefont{Challacombe}},
  \emph{\bibinfo{title}{Linear scaling computation of the Fock matrix. VIII.
  Periodic boundaries for exact exchange at the $\Gamma$-point}}
  (\bibinfo{year}{2004}), \bibinfo{note}{submitted to Phys. Rev. B}.

\bibitem[{\citenamefont{White} \emph{et~al.}(1994)\citenamefont{White, Johnson,
  Gill, and \mbox{Head-Gordon}}}]{CWhite94B}
\bibinfo{author}{\bibfnamefont{C.~A.} \bibnamefont{White}},
  \bibinfo{author}{\bibfnamefont{B.}~\bibnamefont{Johnson}},
  \bibinfo{author}{\bibfnamefont{P.}~\bibnamefont{Gill}}, \bibnamefont{and}
  \bibinfo{author}{\bibfnamefont{M.}~\bibnamefont{\mbox{Head-Gordon}}},
  \bibinfo{journal}{Chem. Phys. Lett.} \textbf{\bibinfo{volume}{230}},
  \bibinfo{pages}{8} (\bibinfo{year}{1994}).

\bibitem[{\citenamefont{White} \emph{et~al.}(1996)\citenamefont{White, Johnson,
  Gill, and \mbox{Head-Gordon}}}]{CWhite96A}
\bibinfo{author}{\bibfnamefont{C.~A.} \bibnamefont{White}},
  \bibinfo{author}{\bibfnamefont{B.~G.} \bibnamefont{Johnson}},
  \bibinfo{author}{\bibfnamefont{P.~M.~W.} \bibnamefont{Gill}},
  \bibnamefont{and}
  \bibinfo{author}{\bibfnamefont{M.}~\bibnamefont{\mbox{Head-Gordon}}},
  \bibinfo{journal}{Chem. Phys. Lett.} \textbf{\bibinfo{volume}{253}},
  \bibinfo{pages}{268} (\bibinfo{year}{1996}).

\bibitem[{\citenamefont{Challacombe}
  \emph{et~al.}(1996{\natexlab{a}})\citenamefont{Challacombe, Schwegler, and
  Alml{\"o}f}}]{MChallacombe96}
\bibinfo{author}{\bibfnamefont{M.}~\bibnamefont{Challacombe}},
  \bibinfo{author}{\bibfnamefont{E.}~\bibnamefont{Schwegler}},
  \bibnamefont{and}
  \bibinfo{author}{\bibfnamefont{J.}~\bibnamefont{Alml{\"o}f}},
  \emph{\bibinfo{title}{Computational Chemistry: Review of Current Trends}}
  (\bibinfo{publisher}{World Scientific}, \bibinfo{address}{Singapore},
  \bibinfo{year}{1996}{\natexlab{a}}), pp. \bibinfo{pages}{53--107}.

\bibitem[{\citenamefont{Challacombe}
  \emph{et~al.}(1996{\natexlab{b}})\citenamefont{Challacombe, Schwegler, and
  Alml{\"o}f}}]{MChallacombe96B}
\bibinfo{author}{\bibfnamefont{M.}~\bibnamefont{Challacombe}},
  \bibinfo{author}{\bibfnamefont{E.}~\bibnamefont{Schwegler}},
  \bibnamefont{and}
  \bibinfo{author}{\bibfnamefont{J.}~\bibnamefont{Alml{\"o}f}},
  \bibinfo{journal}{J. Chem. Phys.} \textbf{\bibinfo{volume}{104}},
  \bibinfo{pages}{4685} (\bibinfo{year}{1996}{\natexlab{b}}).

\bibitem[{\citenamefont{Strain} \emph{et~al.}(1996)\citenamefont{Strain,
  Scuseria, and Frisch}}]{MStrain96}
\bibinfo{author}{\bibfnamefont{M.~C.} \bibnamefont{Strain}},
  \bibinfo{author}{\bibfnamefont{G.~E.} \bibnamefont{Scuseria}},
  \bibnamefont{and} \bibinfo{author}{\bibfnamefont{M.~J.}
  \bibnamefont{Frisch}}, \bibinfo{journal}{Science}
  \textbf{\bibinfo{volume}{271}}, \bibinfo{pages}{51} (\bibinfo{year}{1996}).

\bibitem[{\citenamefont{P{\'e}rez-Jord{\'a} and Yang}(1997)}]{JPerezjorda97}
\bibinfo{author}{\bibfnamefont{J.~M.} \bibnamefont{P{\'e}rez-Jord{\'a}}}
  \bibnamefont{and} \bibinfo{author}{\bibfnamefont{W.~T.} \bibnamefont{Yang}},
  \bibinfo{journal}{J. Chem. Phys.} \textbf{\bibinfo{volume}{107}},
  \bibinfo{pages}{1218} (\bibinfo{year}{1997}).

\bibitem[{\citenamefont{Challacombe and Schwegler}(1997)}]{MChallacombe97}
\bibinfo{author}{\bibfnamefont{M.}~\bibnamefont{Challacombe}} \bibnamefont{and}
  \bibinfo{author}{\bibfnamefont{E.}~\bibnamefont{Schwegler}},
  \bibinfo{journal}{J. Chem. Phys.} \textbf{\bibinfo{volume}{106}},
  \bibinfo{pages}{5526} (\bibinfo{year}{1997}).

\bibitem[{\citenamefont{Tymczak and Challacombe}(2004)}]{CTymczak04a}
\bibinfo{author}{\bibfnamefont{C.~J.} \bibnamefont{Tymczak}} \bibnamefont{and}
  \bibinfo{author}{\bibfnamefont{M.}~\bibnamefont{Challacombe}},
  \emph{\bibinfo{title}{Linear scaling computation of the Fock matrix. VII.
  Periodic Density Functional Theory at the $\Gamma$-point}}
  (\bibinfo{year}{2004}), \bibinfo{note}{submitted to Phys. Rev. B}.

\bibitem[{\citenamefont{P{\'e}rez-Jord{\'a} and Yang}(1995)}]{Jorda95}
\bibinfo{author}{\bibfnamefont{J.~M.} \bibnamefont{P{\'e}rez-Jord{\'a}}}
  \bibnamefont{and} \bibinfo{author}{\bibfnamefont{W.}~\bibnamefont{Yang}},
  \bibinfo{journal}{Chem. Phys. Lett.} \textbf{\bibinfo{volume}{241}},
  \bibinfo{pages}{469} (\bibinfo{year}{1995}).

\bibitem[{\citenamefont{Stratmann} \emph{et~al.}(1996)\citenamefont{Stratmann,
  Scuseria, and Frisch}}]{RStratmann96}
\bibinfo{author}{\bibfnamefont{R.~E.} \bibnamefont{Stratmann}},
  \bibinfo{author}{\bibfnamefont{G.~E.} \bibnamefont{Scuseria}},
  \bibnamefont{and} \bibinfo{author}{\bibfnamefont{M.~J.}
  \bibnamefont{Frisch}}, \bibinfo{journal}{Chem. Phys. Lett.}
  \textbf{\bibinfo{volume}{257}}, \bibinfo{pages}{213} (\bibinfo{year}{1996}).

\bibitem[{\citenamefont{Guerra} \emph{et~al.}(1998)\citenamefont{Guerra,
  Snijders, teVelde, and Baerends}}]{CGuerra98}
\bibinfo{author}{\bibfnamefont{C.~F.} \bibnamefont{Guerra}},
  \bibinfo{author}{\bibfnamefont{J.~G.} \bibnamefont{Snijders}},
  \bibinfo{author}{\bibfnamefont{G.}~\bibnamefont{teVelde}}, \bibnamefont{and}
  \bibinfo{author}{\bibfnamefont{E.~J.} \bibnamefont{Baerends}},
  \bibinfo{journal}{Theor. Chem. Acc.} \textbf{\bibinfo{volume}{99}},
  \bibinfo{pages}{391} (\bibinfo{year}{1998}).

\bibitem[{\citenamefont{Challacombe}(2000)}]{MChallacombe00A}
\bibinfo{author}{\bibfnamefont{M.}~\bibnamefont{Challacombe}},
  \bibinfo{journal}{J. Chem. Phys.} \textbf{\bibinfo{volume}{113}},
  \bibinfo{pages}{10037} (\bibinfo{year}{2000}).

\bibitem[{\citenamefont{Li} \emph{et~al.}(1993)\citenamefont{Li, Nunes, and
  Vanderbilt}}]{XLi93}
\bibinfo{author}{\bibfnamefont{X.~P.} \bibnamefont{Li}},
  \bibinfo{author}{\bibfnamefont{R.~W.} \bibnamefont{Nunes}}, \bibnamefont{and}
  \bibinfo{author}{\bibfnamefont{D.}~\bibnamefont{Vanderbilt}},
  \bibinfo{journal}{Phys. Rev. B} \textbf{\bibinfo{volume}{47}},
  \bibinfo{pages}{10891} (\bibinfo{year}{1993}).

\bibitem[{\citenamefont{Daw}(1993)}]{MDaw93}
\bibinfo{author}{\bibfnamefont{M.~S.} \bibnamefont{Daw}},
  \bibinfo{journal}{Phys. Rev. B} \textbf{\bibinfo{volume}{47}},
  \bibinfo{pages}{10895} (\bibinfo{year}{1993}).

\bibitem[{\citenamefont{Daniels} \emph{et~al.}(1997)\citenamefont{Daniels,
  Millam, and Scuseria}}]{ADaniels97}
\bibinfo{author}{\bibfnamefont{A.~D.} \bibnamefont{Daniels}},
  \bibinfo{author}{\bibfnamefont{J.~M.} \bibnamefont{Millam}},
  \bibnamefont{and} \bibinfo{author}{\bibfnamefont{G.~E.}
  \bibnamefont{Scuseria}}, \bibinfo{journal}{J. Chem. Phys.}
  \textbf{\bibinfo{volume}{107}}, \bibinfo{pages}{425} (\bibinfo{year}{1997}).

\bibitem[{\citenamefont{Palser and Manolopoulos}(1998)}]{APalser98}
\bibinfo{author}{\bibfnamefont{A.~H.~R.} \bibnamefont{Palser}}
  \bibnamefont{and} \bibinfo{author}{\bibfnamefont{D.~E.}
  \bibnamefont{Manolopoulos}}, \bibinfo{journal}{Phys. Rev. B}
  \textbf{\bibinfo{volume}{58}}, \bibinfo{pages}{12704} (\bibinfo{year}{1998}).

\bibitem[{\citenamefont{Challacombe}(1999)}]{MChallacombe99}
\bibinfo{author}{\bibfnamefont{M.}~\bibnamefont{Challacombe}},
  \bibinfo{journal}{J. Chem. Phys.} \textbf{\bibinfo{volume}{110}},
  \bibinfo{pages}{2332} (\bibinfo{year}{1999}).

\bibitem[{\citenamefont{Niklasson}(2002)}]{ANiklasson02A}
\bibinfo{author}{\bibfnamefont{A.~M.~N.} \bibnamefont{Niklasson}},
  \bibinfo{journal}{Phys. Rev. B} \textbf{\bibinfo{volume}{66}},
  \bibinfo{pages}{155115} (\bibinfo{year}{2002}).

\bibitem[{\citenamefont{Niklasson} \emph{et~al.}(2003)\citenamefont{Niklasson,
  Tymczak, and Challacombe}}]{ANiklasson03}
\bibinfo{author}{\bibfnamefont{A.~M.~N.} \bibnamefont{Niklasson}},
  \bibinfo{author}{\bibfnamefont{C.~J.} \bibnamefont{Tymczak}},
  \bibnamefont{and}
  \bibinfo{author}{\bibfnamefont{M.}~\bibnamefont{Challacombe}},
  \bibinfo{journal}{J. Chem. Phys.}
  \textbf{\bibinfo{volume}{118}}(\bibinfo{number}{19}), \bibinfo{pages}{8611}
  (\bibinfo{year}{2003}).

\bibitem[{\citenamefont{Schmidt} \emph{et~al.}(1993)\citenamefont{Schmidt,
  Baldridge, Boatz, Elbert, Gordon, Jensen, Koseki, Matsunaga, Nguyen, Su,
  Windus, Dupuis} \emph{et~al.}}]{MSchmidt93}
\bibinfo{author}{\bibfnamefont{M.~W.} \bibnamefont{Schmidt}},
  \bibinfo{author}{\bibfnamefont{K.~K.} \bibnamefont{Baldridge}},
  \bibinfo{author}{\bibfnamefont{J.~A.} \bibnamefont{Boatz}},
  \bibinfo{author}{\bibfnamefont{S.~T.} \bibnamefont{Elbert}},
  \bibinfo{author}{\bibfnamefont{M.~S.} \bibnamefont{Gordon}},
  \bibinfo{author}{\bibfnamefont{J.~H.} \bibnamefont{Jensen}},
  \bibinfo{author}{\bibfnamefont{S.}~\bibnamefont{Koseki}},
  \bibinfo{author}{\bibfnamefont{N.}~\bibnamefont{Matsunaga}},
  \bibinfo{author}{\bibfnamefont{K.~A.} \bibnamefont{Nguyen}},
  \bibinfo{author}{\bibfnamefont{S.~J.} \bibnamefont{Su}},
  \bibinfo{author}{\bibfnamefont{T.~L.} \bibnamefont{Windus}},
  \bibinfo{author}{\bibfnamefont{M.}~\bibnamefont{Dupuis}}, \emph{et~al.},
  \bibinfo{journal}{J. Comput. Chem.} \textbf{\bibinfo{volume}{14}},
  \bibinfo{pages}{1347} (\bibinfo{year}{1993}).

\bibitem[{\citenamefont{Colvin} \emph{et~al.}(1993)\citenamefont{Colvin,
  Janssen, Whiteside, and Tong}}]{MColvin93}
\bibinfo{author}{\bibfnamefont{M.~E.} \bibnamefont{Colvin}},
  \bibinfo{author}{\bibfnamefont{C.~L.} \bibnamefont{Janssen}},
  \bibinfo{author}{\bibfnamefont{R.~A.} \bibnamefont{Whiteside}},
  \bibnamefont{and} \bibinfo{author}{\bibfnamefont{C.~H.} \bibnamefont{Tong}},
  \bibinfo{journal}{Theor. Chim. Acta} \textbf{\bibinfo{volume}{84}},
  \bibinfo{pages}{301} (\bibinfo{year}{1993}).

\bibitem[{\citenamefont{Harrison and Shepard}(1994)}]{Harrison_94v45}
\bibinfo{author}{\bibfnamefont{R.~J.} \bibnamefont{Harrison}} \bibnamefont{and}
  \bibinfo{author}{\bibfnamefont{R.}~\bibnamefont{Shepard}},
  \bibinfo{journal}{Annu. Rev. Phys. Chem.} \textbf{\bibinfo{volume}{45}},
  \bibinfo{pages}{623} (\bibinfo{year}{1994}).

\bibitem[{\citenamefont{Guerra} \emph{et~al.}(1995)\citenamefont{Guerra,
  Visser, Snijders, te~Velde, and Baerends}}]{Guerra_95}
\bibinfo{author}{\bibfnamefont{C.~F.} \bibnamefont{Guerra}},
  \bibinfo{author}{\bibfnamefont{O.}~\bibnamefont{Visser}},
  \bibinfo{author}{\bibfnamefont{J.~G.} \bibnamefont{Snijders}},
  \bibinfo{author}{\bibfnamefont{G.}~\bibnamefont{te~Velde}}, \bibnamefont{and}
  \bibinfo{author}{\bibfnamefont{E.~J.} \bibnamefont{Baerends}}, in
  \emph{\bibinfo{booktitle}{Methods and techniques for Computational
  Chemistry}}, edited by
  \bibinfo{editor}{\bibfnamefont{E.}~\bibnamefont{Clementi}} \bibnamefont{and}
  \bibinfo{editor}{\bibfnamefont{G.}~\bibnamefont{Corongiu}}
  (\bibinfo{publisher}{STEF, Cagliari}, \bibinfo{year}{1995}), p.
  \bibinfo{pages}{305}.

\bibitem[{\citenamefont{Furlani and King}(1995)}]{TFurlani95}
\bibinfo{author}{\bibfnamefont{T.~R.} \bibnamefont{Furlani}} \bibnamefont{and}
  \bibinfo{author}{\bibfnamefont{H.~F.} \bibnamefont{King}},
  \bibinfo{journal}{J. Comput. Chem.} \textbf{\bibinfo{volume}{16}},
  \bibinfo{pages}{91} (\bibinfo{year}{1995}).

\bibitem[{\citenamefont{Sosa} \emph{et~al.}(1998)\citenamefont{Sosa, Ochterski,
  Carpenter, and Frisch}}]{Sosa_98v19}
\bibinfo{author}{\bibfnamefont{C.~P.} \bibnamefont{Sosa}},
  \bibinfo{author}{\bibfnamefont{J.}~\bibnamefont{Ochterski}},
  \bibinfo{author}{\bibfnamefont{J.}~\bibnamefont{Carpenter}},
  \bibnamefont{and} \bibinfo{author}{\bibfnamefont{M.~J.}
  \bibnamefont{Frisch}}, \bibinfo{journal}{J. Comput. Chem.}
  \textbf{\bibinfo{volume}{19}}, \bibinfo{pages}{1053} (\bibinfo{year}{1998}).

\bibitem[{\citenamefont{von Arnim and Ahlrichs}(1998)}]{vonArnim98}
\bibinfo{author}{\bibfnamefont{M.}~\bibnamefont{von Arnim}} \bibnamefont{and}
  \bibinfo{author}{\bibfnamefont{R.}~\bibnamefont{Ahlrichs}},
  \bibinfo{journal}{J. Comput. Chem.} \textbf{\bibinfo{volume}{19}},
  \bibinfo{pages}{1746} (\bibinfo{year}{1998}).

\bibitem[{\citenamefont{Furlani} \emph{et~al.}(2000)\citenamefont{Furlani,
  Kong, and Gill}}]{Furlani_00v128}
\bibinfo{author}{\bibfnamefont{T.~R.} \bibnamefont{Furlani}},
  \bibinfo{author}{\bibfnamefont{J.}~\bibnamefont{Kong}}, \bibnamefont{and}
  \bibinfo{author}{\bibfnamefont{P.~M.~W.} \bibnamefont{Gill}},
  \bibinfo{journal}{Comput. Phys. Commun.} \textbf{\bibinfo{volume}{128}},
  \bibinfo{pages}{170} (\bibinfo{year}{2000}).

\bibitem[{\citenamefont{Sosa} \emph{et~al.}(2000)\citenamefont{Sosa, Scalmani,
  Gomperts, and Frisch}}]{Sosa_00v26}
\bibinfo{author}{\bibfnamefont{C.~P.} \bibnamefont{Sosa}},
  \bibinfo{author}{\bibfnamefont{G.}~\bibnamefont{Scalmani}},
  \bibinfo{author}{\bibfnamefont{R.}~\bibnamefont{Gomperts}}, \bibnamefont{and}
  \bibinfo{author}{\bibfnamefont{M.~J.} \bibnamefont{Frisch}},
  \bibinfo{journal}{Parallel Comput.} \textbf{\bibinfo{volume}{26}},
  \bibinfo{pages}{843} (\bibinfo{year}{2000}).

\bibitem[{\citenamefont{Kendall} \emph{et~al.}(2000)\citenamefont{Kendall,
  Apra, Bernholdt, Bylaska, Dupuis, Fann, Harrison, Ju, Nichols, Nieplocha,
  Straatsma, Windus} \emph{et~al.}}]{RKendall00}
\bibinfo{author}{\bibfnamefont{R.~A.} \bibnamefont{Kendall}},
  \bibinfo{author}{\bibfnamefont{E.}~\bibnamefont{Apra}},
  \bibinfo{author}{\bibfnamefont{D.~E.} \bibnamefont{Bernholdt}},
  \bibinfo{author}{\bibfnamefont{E.~J.} \bibnamefont{Bylaska}},
  \bibinfo{author}{\bibfnamefont{M.}~\bibnamefont{Dupuis}},
  \bibinfo{author}{\bibfnamefont{G.~I.} \bibnamefont{Fann}},
  \bibinfo{author}{\bibfnamefont{R.~J.} \bibnamefont{Harrison}},
  \bibinfo{author}{\bibfnamefont{J.}~\bibnamefont{Ju}},
  \bibinfo{author}{\bibfnamefont{J.~A.} \bibnamefont{Nichols}},
  \bibinfo{author}{\bibfnamefont{J.}~\bibnamefont{Nieplocha}},
  \bibinfo{author}{\bibfnamefont{T.~P.} \bibnamefont{Straatsma}},
  \bibinfo{author}{\bibfnamefont{T.~L.} \bibnamefont{Windus}}, \emph{et~al.},
  \bibinfo{journal}{Comput. Phys. Commun.} \textbf{\bibinfo{volume}{128}},
  \bibinfo{pages}{260} (\bibinfo{year}{2000}).

\bibitem[{\citenamefont{Fletcher} \emph{et~al.}(2000)\citenamefont{Fletcher,
  Schmidt, Bode, and Gordon}}]{GFletcher00}
\bibinfo{author}{\bibfnamefont{G.~D.} \bibnamefont{Fletcher}},
  \bibinfo{author}{\bibfnamefont{M.~W.} \bibnamefont{Schmidt}},
  \bibinfo{author}{\bibfnamefont{B.~M.} \bibnamefont{Bode}}, \bibnamefont{and}
  \bibinfo{author}{\bibfnamefont{M.~S.} \bibnamefont{Gordon}},
  \bibinfo{journal}{Comput. Phys. Commun.} \textbf{\bibinfo{volume}{128}},
  \bibinfo{pages}{190} (\bibinfo{year}{2000}).

\bibitem[{\citenamefont{Baker and Pulay}(2002)}]{Baker_02v23}
\bibinfo{author}{\bibfnamefont{J.}~\bibnamefont{Baker}} \bibnamefont{and}
  \bibinfo{author}{\bibfnamefont{P.}~\bibnamefont{Pulay}}, \bibinfo{journal}{J.
  Comput. Chem.} \textbf{\bibinfo{volume}{23}}, \bibinfo{pages}{1150}
  (\bibinfo{year}{2002}).

\bibitem[{\citenamefont{Baker} \emph{et~al.}(2004)\citenamefont{Baker,
  F\"{u}sti-Molnar, and Pulay}}]{JBaker04}
\bibinfo{author}{\bibfnamefont{J.}~\bibnamefont{Baker}},
  \bibinfo{author}{\bibfnamefont{L.}~\bibnamefont{F\"{u}sti-Molnar}},
  \bibnamefont{and} \bibinfo{author}{\bibfnamefont{P.}~\bibnamefont{Pulay}},
  \bibinfo{journal}{J. Phys. Chem. A} \textbf{\bibinfo{volume}{108}},
  \bibinfo{pages}{3040} (\bibinfo{year}{2004}).

\bibitem[{\citenamefont{Gan and Challacombe}(2003)}]{CGan03}
\bibinfo{author}{\bibfnamefont{C.~K.} \bibnamefont{Gan}} \bibnamefont{and}
  \bibinfo{author}{\bibfnamefont{M.}~\bibnamefont{Challacombe}},
  \bibinfo{journal}{J. Chem. Phys.} \textbf{\bibinfo{volume}{118}},
  \bibinfo{pages}{9128} (\bibinfo{year}{2003}).

\bibitem[{\citenamefont{Barnes and Hut}(1986)}]{JBarnes86}
\bibinfo{author}{\bibfnamefont{J.}~\bibnamefont{Barnes}} \bibnamefont{and}
  \bibinfo{author}{\bibfnamefont{P.}~\bibnamefont{Hut}},
  \bibinfo{journal}{Nature} \textbf{\bibinfo{volume}{324}},
  \bibinfo{pages}{446} (\bibinfo{year}{1986}).

\bibitem[{\citenamefont{Greengard and Rokhlin}(1987)}]{LGreengard87}
\bibinfo{author}{\bibfnamefont{L.}~\bibnamefont{Greengard}} \bibnamefont{and}
  \bibinfo{author}{\bibfnamefont{V.}~\bibnamefont{Rokhlin}},
  \bibinfo{journal}{J. Comp. Phys.} \textbf{\bibinfo{volume}{73}},
  \bibinfo{pages}{325} (\bibinfo{year}{1987}).

\bibitem[{\citenamefont{Anderson}(1992)}]{CRAnderson92}
\bibinfo{author}{\bibfnamefont{C.~R.} \bibnamefont{Anderson}},
  \bibinfo{journal}{SIAM J. Sci. Stat. Comput.} \textbf{\bibinfo{volume}{13}},
  \bibinfo{pages}{923} (\bibinfo{year}{1992}).

\bibitem[{\citenamefont{Warren and Salmon}(1992)}]{MWarren92}
\bibinfo{author}{\bibfnamefont{M.~S.} \bibnamefont{Warren}} \bibnamefont{and}
  \bibinfo{author}{\bibfnamefont{J.~K.} \bibnamefont{Salmon}},
  \bibinfo{journal}{Proceedings of Supercomputing '92} p. \bibinfo{pages}{570}
  (\bibinfo{year}{1992}).

\bibitem[{\citenamefont{Grama} \emph{et~al.}(1994)\citenamefont{Grama, Kumar,
  and Sameh}}]{AGrama94}
\bibinfo{author}{\bibfnamefont{A.~Y.} \bibnamefont{Grama}},
  \bibinfo{author}{\bibfnamefont{V.}~\bibnamefont{Kumar}}, \bibnamefont{and}
  \bibinfo{author}{\bibfnamefont{A.}~\bibnamefont{Sameh}},
  \bibinfo{journal}{Proceedings of Supercomputing '94} p. \bibinfo{pages}{439}
  (\bibinfo{year}{1994}).

\bibitem[{\citenamefont{Warren and Salmon}(1995)}]{MWarren95b}
\bibinfo{author}{\bibfnamefont{M.~S.} \bibnamefont{Warren}} \bibnamefont{and}
  \bibinfo{author}{\bibfnamefont{J.~K.} \bibnamefont{Salmon}},
  \bibinfo{journal}{Comput. Phys. Commun.} \textbf{\bibinfo{volume}{87}},
  \bibinfo{pages}{266} (\bibinfo{year}{1995}).

\bibitem[{\citenamefont{Singh} \emph{et~al.}(1993)\citenamefont{Singh, Holt,
  Hennessy, and Gupta}}]{Singh93}
\bibinfo{author}{\bibfnamefont{J.~P.} \bibnamefont{Singh}},
  \bibinfo{author}{\bibfnamefont{C.}~\bibnamefont{Holt}},
  \bibinfo{author}{\bibfnamefont{J.~L.} \bibnamefont{Hennessy}},
  \bibnamefont{and} \bibinfo{author}{\bibfnamefont{A.}~\bibnamefont{Gupta}},
  \bibinfo{journal}{Proceedings of Supercomputing '93} p.~\bibinfo{pages}{54}
  (\bibinfo{year}{1993}).

\bibitem[{\citenamefont{Singh} \emph{et~al.}(1995)\citenamefont{Singh, Holt,
  Totsuka, Gupta, and Hennessy}}]{Singh_95v27}
\bibinfo{author}{\bibfnamefont{J.~P.} \bibnamefont{Singh}},
  \bibinfo{author}{\bibfnamefont{C.}~\bibnamefont{Holt}},
  \bibinfo{author}{\bibfnamefont{T.}~\bibnamefont{Totsuka}},
  \bibinfo{author}{\bibfnamefont{A.}~\bibnamefont{Gupta}}, \bibnamefont{and}
  \bibinfo{author}{\bibfnamefont{J.}~\bibnamefont{Hennessy}},
  \bibinfo{journal}{J. Para. Distr. Comput.} \textbf{\bibinfo{volume}{27}},
  \bibinfo{pages}{118} (\bibinfo{year}{1995}).

\bibitem[{\citenamefont{Hu and Johnsson}(1996)}]{YHu96}
\bibinfo{author}{\bibfnamefont{Y.}~\bibnamefont{Hu}} \bibnamefont{and}
  \bibinfo{author}{\bibfnamefont{S.~L.} \bibnamefont{Johnsson}},
  \bibinfo{journal}{Int. J. Supercomput. Appl. High Perfom. Comput.}
  \textbf{\bibinfo{volume}{10}}, \bibinfo{pages}{3} (\bibinfo{year}{1996}).

\bibitem[{\citenamefont{Grama} \emph{et~al.}(1998)\citenamefont{Grama, Kumar,
  and Sameh}}]{Grama_98v24}
\bibinfo{author}{\bibfnamefont{A.}~\bibnamefont{Grama}},
  \bibinfo{author}{\bibfnamefont{V.}~\bibnamefont{Kumar}}, \bibnamefont{and}
  \bibinfo{author}{\bibfnamefont{A.}~\bibnamefont{Sameh}},
  \bibinfo{journal}{Parallel Comput.} \textbf{\bibinfo{volume}{24}},
  \bibinfo{pages}{797} (\bibinfo{year}{1998}).

\bibitem[{\citenamefont{Gibbon and Sutmann}(2002)}]{PGibbon02}
\bibinfo{author}{\bibfnamefont{P.}~\bibnamefont{Gibbon}} \bibnamefont{and}
  \bibinfo{author}{\bibfnamefont{G.}~\bibnamefont{Sutmann}},
  \emph{\bibinfo{title}{Long-Range Interactions in Many-Particle Simulation}}
  (\bibinfo{publisher}{John Wiley and Sons}, \bibinfo{address}{John von Neumann
  Institute for Computing, Julich}, \bibinfo{year}{2002}), pp.
  \bibinfo{pages}{467--506}.

\bibitem[{\citenamefont{Antonuccio-Delogu}
  \emph{et~al.}(2003)\citenamefont{Antonuccio-Delogu, Becciani, and
  Ferro}}]{Antonuccio-Delogu03}
\bibinfo{author}{\bibfnamefont{V.}~\bibnamefont{Antonuccio-Delogu}},
  \bibinfo{author}{\bibfnamefont{U.}~\bibnamefont{Becciani}}, \bibnamefont{and}
  \bibinfo{author}{\bibfnamefont{D.}~\bibnamefont{Ferro}},
  \bibinfo{journal}{Comput. Phys. Commun.} \textbf{\bibinfo{volume}{155}},
  \bibinfo{pages}{159} (\bibinfo{year}{2003}).

\bibitem[{\citenamefont{Wilkinson and Allen}(1999)}]{BWilkinson99}
\bibinfo{author}{\bibfnamefont{B.}~\bibnamefont{Wilkinson}} \bibnamefont{and}
  \bibinfo{author}{\bibfnamefont{M.}~\bibnamefont{Allen}},
  \emph{\bibinfo{title}{Parallel Programming}} (\bibinfo{publisher}{Prentice
  Hall}, \bibinfo{address}{Upper Saddle River, NJ}, \bibinfo{year}{1999}).

\bibitem[{\citenamefont{Pilkington and Baden}(1996)}]{JPilkington96}
\bibinfo{author}{\bibfnamefont{J.~R.} \bibnamefont{Pilkington}}
  \bibnamefont{and} \bibinfo{author}{\bibfnamefont{S.~B.} \bibnamefont{Baden}},
  \bibinfo{journal}{IEEE Trans. Para. Distr. Sys.}
  \textbf{\bibinfo{volume}{7}}, \bibinfo{pages}{288} (\bibinfo{year}{1996}).

\bibitem[{\citenamefont{Warren and Salmon}(1993)}]{MWarren93}
\bibinfo{author}{\bibfnamefont{M.~S.} \bibnamefont{Warren}} \bibnamefont{and}
  \bibinfo{author}{\bibfnamefont{J.~K.} \bibnamefont{Salmon}},
  \bibinfo{journal}{Proceedings of Supercomputing '93} p.~\bibinfo{pages}{12}
  (\bibinfo{year}{1993}).

\bibitem[{\citenamefont{McMurchie and Davidson}(1978)}]{LMcmurchie78}
\bibinfo{author}{\bibfnamefont{L.~E.} \bibnamefont{McMurchie}}
  \bibnamefont{and} \bibinfo{author}{\bibfnamefont{E.~R.}
  \bibnamefont{Davidson}}, \bibinfo{journal}{J. Comp. Phys.}
  \textbf{\bibinfo{volume}{26}}, \bibinfo{pages}{218} (\bibinfo{year}{1978}).

\bibitem[{\citenamefont{Bentley and Friedman}(1979)}]{Bentley79}
\bibinfo{author}{\bibfnamefont{J.~L.} \bibnamefont{Bentley}} \bibnamefont{and}
  \bibinfo{author}{\bibfnamefont{J.~H.} \bibnamefont{Friedman}},
  \bibinfo{journal}{ACM Comp. Surv.} \textbf{\bibinfo{volume}{11}},
  \bibinfo{pages}{397} (\bibinfo{year}{1979}).

\bibitem[{\citenamefont{Bentley}(1980)}]{Bentley80}
\bibinfo{author}{\bibfnamefont{J.~L.} \bibnamefont{Bentley}},
  \bibinfo{journal}{Commun. ACM} \textbf{\bibinfo{volume}{23}},
  \bibinfo{pages}{214} (\bibinfo{year}{1980}).

\bibitem[{\citenamefont{Gaede and G{\"u}nther}(1998)}]{Gaede98}
\bibinfo{author}{\bibfnamefont{V.}~\bibnamefont{Gaede}} \bibnamefont{and}
  \bibinfo{author}{\bibfnamefont{O.}~\bibnamefont{G{\"u}nther}},
  \bibinfo{journal}{ACM Comput. Surv.} \textbf{\bibinfo{volume}{30}},
  \bibinfo{pages}{170} (\bibinfo{year}{1998}).

\bibitem[{\citenamefont{Ahmadi and Alml{\"o}f}(1995)}]{GAhmadi95}
\bibinfo{author}{\bibfnamefont{G.~R.} \bibnamefont{Ahmadi}} \bibnamefont{and}
  \bibinfo{author}{\bibfnamefont{J.}~\bibnamefont{Alml{\"o}f}},
  \bibinfo{journal}{Chem. Phys. Lett.} \textbf{\bibinfo{volume}{246}},
  \bibinfo{pages}{364} (\bibinfo{year}{1995}).

\bibitem[{\citenamefont{Challacombe}
  \emph{et~al.}(1997)\citenamefont{Challacombe, White, and
  Head-Gordon}}]{MChallacombe97D}
\bibinfo{author}{\bibfnamefont{M.}~\bibnamefont{Challacombe}},
  \bibinfo{author}{\bibfnamefont{C.}~\bibnamefont{White}}, \bibnamefont{and}
  \bibinfo{author}{\bibfnamefont{M.}~\bibnamefont{Head-Gordon}},
  \bibinfo{journal}{J. Chem. Phys.} \textbf{\bibinfo{volume}{107}},
  \bibinfo{pages}{10131} (\bibinfo{year}{1997}).

\bibitem[{mpi(1998)}]{mpi}
\emph{\bibinfo{title}{Message Passing Interface Forum. MPI: A message-passing
  interface standard (version 2.0)}} (\bibinfo{year}{1998}),
  \bibinfo{note}{{\tt http://www.mpi-forum.org}}.

\bibitem[{\citenamefont{Press} \emph{et~al.}(1992)\citenamefont{Press,
  Teukolsky, Vetterling, and Flannery}}]{WPress92}
\bibinfo{author}{\bibfnamefont{W.~H.} \bibnamefont{Press}},
  \bibinfo{author}{\bibfnamefont{S.~A.} \bibnamefont{Teukolsky}},
  \bibinfo{author}{\bibfnamefont{W.~T.} \bibnamefont{Vetterling}},
  \bibnamefont{and} \bibinfo{author}{\bibfnamefont{B.~P.}
  \bibnamefont{Flannery}}, \emph{\bibinfo{title}{Numerical Recipies in
  \mbox{FORTRAN}}} (\bibinfo{publisher}{Cambridge University Press},
  \bibinfo{address}{Port Chester, NY}, \bibinfo{year}{1992}).

\bibitem[{\citenamefont{Challacombe}
  \emph{et~al.}(2001)\citenamefont{Challacombe, Schwegler, Tymczak, Gan,
  Nemeth, Weber, Niklasson, and Henkelman}}]{MondoSCF}
\bibinfo{author}{\bibfnamefont{M.}~\bibnamefont{Challacombe}},
  \bibinfo{author}{\bibfnamefont{E.}~\bibnamefont{Schwegler}},
  \bibinfo{author}{\bibfnamefont{C.~J.} \bibnamefont{Tymczak}},
  \bibinfo{author}{\bibfnamefont{C.~K.} \bibnamefont{Gan}},
  \bibinfo{author}{\bibfnamefont{K.}~\bibnamefont{Nemeth}},
  \bibinfo{author}{\bibfnamefont{V.}~\bibnamefont{Weber}},
  \bibinfo{author}{\bibfnamefont{A.~M.~N.} \bibnamefont{Niklasson}},
  \bibnamefont{and}
  \bibinfo{author}{\bibfnamefont{G.}~\bibnamefont{Henkelman}},
  \emph{\bibinfo{title}{{\sc MondoSCF} v1.0$\alpha$9, A program suite for
  massively parallel, linear scaling SCF theory and {\it ab initio} molecular
  dynamics.}} (\bibinfo{year}{2001}),
  \urlprefix\url{http://www.t12.lanl.gov/home/mchalla/},
  \bibinfo{note}{\mbox{L}os Alamos National Laboratory (LA-CC 01-2), Copyright
  University of California.}

\bibitem[{\citenamefont{Gan}
  \emph{et~al.}(2004{\natexlab{a}})\citenamefont{Gan, Sewell, and
  Challacombe}}]{CGan04A}
\bibinfo{author}{\bibfnamefont{C.~K.} \bibnamefont{Gan}},
  \bibinfo{author}{\bibfnamefont{T.~D.} \bibnamefont{Sewell}},
  \bibnamefont{and}
  \bibinfo{author}{\bibfnamefont{M.}~\bibnamefont{Challacombe}},
  \bibinfo{journal}{Phys. Rev. B} \textbf{\bibinfo{volume}{69}},
  \bibinfo{pages}{035116} (\bibinfo{year}{2004}{\natexlab{a}}).

\bibitem[{\citenamefont{Lewis} \emph{et~al.}(2000)\citenamefont{Lewis, Sewell,
  Evans, and Voth}}]{JPLewis00}
\bibinfo{author}{\bibfnamefont{J.~P.} \bibnamefont{Lewis}},
  \bibinfo{author}{\bibfnamefont{T.~D.} \bibnamefont{Sewell}},
  \bibinfo{author}{\bibfnamefont{R.~B.} \bibnamefont{Evans}}, \bibnamefont{and}
  \bibinfo{author}{\bibfnamefont{G.~A.} \bibnamefont{Voth}},
  \bibinfo{journal}{J. Phys. Chem. B} \textbf{\bibinfo{volume}{104}},
  \bibinfo{pages}{1009} (\bibinfo{year}{2000}).

\bibitem[{\citenamefont{Gan}
  \emph{et~al.}(2004{\natexlab{b}})\citenamefont{Gan, Sewell, and
  Challacombe}}]{CGan04C}
\bibinfo{author}{\bibfnamefont{C.~K.} \bibnamefont{Gan}},
  \bibinfo{author}{\bibfnamefont{T.~D.} \bibnamefont{Sewell}},
  \bibnamefont{and}
  \bibinfo{author}{\bibfnamefont{M.}~\bibnamefont{Challacombe}},
  \emph{\bibinfo{title}{Equation of state of $\beta$-HMX
  (octahydro-1,3,5,7-tetranitro-1,3,5,7-tetrazocine)}}
  (\bibinfo{year}{2004}{\natexlab{b}}), \bibinfo{note}{in preparation}.

\end{thebibliography}
\end{document}